\documentclass[aps, pre]{revtex4}

\usepackage{graphicx}

\begin{document}

\title{ Entropically-Stabilized Self-compactification in Model Colloidal
Systems }
\author{Juyong Park and Alexei V. Tkachenko}
\affiliation{Department of Physics, University of Michigan, 500 E. University Ave., Ann
Arbor, 48109 MI, USA}

\begin{abstract}
We discuss the phenomenon of spontaneous self--compactification in a model
colloidal system, proposed in a recent work on DNA--mediated self--assembly.
We focus on the effect of thermal fluctuations on the stability of
membrane-like self--assembled phase with in-plane square order.
Surprisingly, the fluctuations are shown to enhance the stability of this
quasi--2D phase with respect to transition to alternative 3D structures.

\end{abstract}
\maketitle

%%\DOI{123}                       % do not fill in
%%\idline{C}{1, 1--11}{1}         % do not fill in
%%\editorial{}{}{}{}              % do not fill in
%

\section{Introduction}

Effects of physical dimensionality on crystallization are among the most important
problems in Condensed Matter Physics. As was established by Landau and
Pierls \cite{LP}, the long-range crystalline order in 2D is universally
destroyed by thermal fluctuations. The problem has been revisited in 1970s
by Kosterlitz and Thouless \cite{LP}-\cite{BKT}, who have shown that
crystals do exist in 2D, within a new topological definition. Nevertheless,
their melting temperature is believed to be always lower than in 3D systems
(assuming the same interparticle potential). One of the manifestations of
this effect is the phenomenon of surface melting: a microscopic liquid layer
normally appear at the interface of a crystalline solid well below its bulk
melting temperature.

In this paper, we describe a remarkable example \ in which the thermal
fluctuation \emph{stabilize} a 2D crystalline solid, embedded in 3D physical
space, with respect to transition to an alternative 3D structure. The model
system discussed below has been introduced in the recent work by one of us 
\cite{AT}, in order to describe DNA--assisted self--assembly of colloids.
Its essential ingredients are cohesive interparticle interactions and
medium--range soft core repulsion. The binary system of same--size spheres
(A and B) discussed in Ref. \cite{AT}, combines the repulsive potential $%
U(r) $ acting between same--type particles, with A--B attraction. As was
shown at that work, both interactions may be induced by properly designed
DNA. It was found that this colloid--DNA mixture may exhibit an unusually
diverse phase diagram as a function of two control parameter: the relative
strength of attraction and repulsion, and aspect ratio $\xi /a$ ($\xi $ is
the range of repulsive potential $U(r)$, and $a$ is the particle diameter).

Among various self--assembled phases expected for that system, it was
especially striking to find quasi-2D membrane with the in--plane square
order (SQ). In other words, according to our calculations, this 3D system
may prefer to self--assemble into a lower--dimensional structure. We will
refer to this phenomenon as spontaneous self--compactification. Of course,
there are other known examples of self--compactified structures in condensed
matter, such as lipid membranes \cite{membr}. However, our case is quite
unique because it is based on \emph{isotropic pair potentials} (in contrast
to anisotropic interactions between lipids, or covalent bonding of carbon
atoms in graphite). Note also that at the found SQ phase is not a stacking
of weakly coupled layers (lamella--like), but rather an isolated
membrane--like structure.

In our early calculations, we have only accounted for the interplay of
repulsive and attractive energies, while the thermal fluctuations were
totally ignored. Even though this approximation is applicable when the
characteristic energies considerably exceed $k_{B}T$, the entropic effects
are expected to be significant in any realistic case. Given that
fluctuations are known to strongly affect 2D crystals, one might wonder
whether the phenomenon of self--compactification will still be present if
the fluctuations are introduced. Below we present the detailed study of this
question.

\section{The model and its generic features}

Before going into a specific example, we describe our model and its generic
features. We start with particles packed into ideal crystalline lattices,
whose fluctuation free energies are to be compared. The interparticle
potentials will be replaced with linear springs, whose spring constants
correspond to the second derivatives of the corresponding potentials. The
first derivatives of the potentials will give rise to a pre--existing
stresses in those springs. We consider only very short--range interactions,
so that any connections beyond second nearest neighbors will be neglected.

\begin{figure}[tbp]
\includegraphics[
height=1.9in,
width=4.5in
]{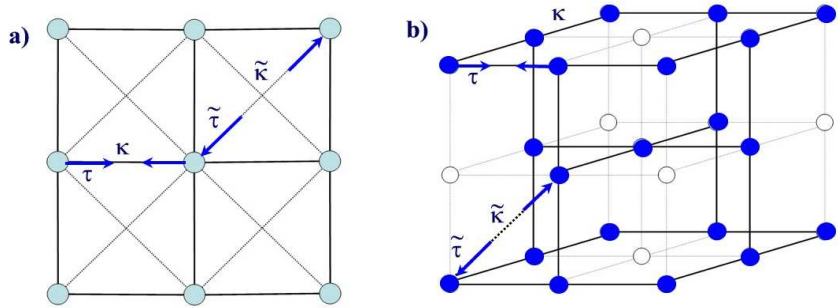}
\caption{SQ (a) and its 3D-dual phase (b).}
\label{SQ}
\end{figure}

Let $\kappa $ and $\tilde{\kappa}$ be spring constants for the first and
second nearest neighbor bonds, respectively. Repulsion between the second
nearest neighbors induces tension $\tilde{\tau}$ in $\tilde{\kappa}$%
--springs, which should be balanced by an appropriate compressive force $%
-\tau $ $\ $in $\kappa $--springs. If the interaction range $\xi $ is much
shorter than nearest neighbors distance $a$, one expects $\kappa \gg \tilde{%
\kappa}$ , and $\tau \sim \tilde{\tau}\sim \tilde{\kappa}/\xi $. \ This
gives rise to a hierarchy of elastic constants in this system: $\kappa \gg 
\tilde{\kappa}$ $\gg \tau /a$. Below, we will use harmonic analysis to
diagonalize the phonon Hamiltonian and find the fluctational contributions
to Free Energies of the competing phases. As an example, we compare 2D
square lattice (SQ) (embedded in 3D physical space) to an alternative
three--dimensional phase with a very similar local structure. Figure \ref{SQ}
shows this 3D counterpart of SQ (referred below as "dual phase"), which also
has four nearest neighbors lying in one plane around each particle. The
difference from SQ is that now there are eight, rather than four second
nearest neighbors for each site. Because of this difference in the number of
repulsive bonds, SQ phase is generally preferred energetically over its 3D
dual, at the zero temperature limit. Below, we will see whether the free
energy balance between the two phases may be reversed by the thermal
fluctuations.

Before we present the exact results for SQ, and its 3D-dual phase, we
discuss certain general features of this problem. Given the hierarchy of the
spring constants, one can distinguish between several kinds of phonon modes.
Namely, there are \emph{hard} modes, which involve deformations of strong $%
\kappa $--springs, and \emph{soft} modes which only depend on spring
constant $\tilde{\kappa}$. There are $3$ degrees of freedom per particle,
and $2$ of them correspond to hard modes, both for 2D lattice and its 3D
dual phase. In 3D structure, the third mode is the soft one ($\tilde{\kappa}$%
-mode). However, if there were no stresses $\tau $ and $\tilde{\tau}$,
within harmonic approximation there would be no restoring force for the
out-of-plane fluctuations in the 2D structure. This means that in 2D lattice
the effective spring constant for the transverse modes should be of the
order of $\tau /a$ (since $k\gg \tilde{\kappa}\gg \tau /a$, we may
justifiably call these modes of 2D lattice $\emph{supersoft}$). \ Because of
the replacement of the soft modes with the supersoft ones, we do expect the
entropy of the quasi--2D structure to be higher, which will further shift
the free energy balance to its favor. In addition, in small--$q$ limit, the
out-of-plane modes in 2D structure will universally have dispersion $\sim
q^{4}$ (associated with bending rigidity). This will give another negative
correction to free energy compared to dispersion $\sim q^{2}$ of regular
acoustic phonons. Note that these arguments are quite universal, and can be
applied beyond the particular case of SQ lattice.

\section{Example: Square vs. its 3D-dual}

We now proceed with the discussion of the specific example, 2D SQ vs. its
3D-dual phase. The latter 3D structure may be obtained from a simple cubic
lattice, by removing the particles occupying sites . $(2m,2k,2l+1)$ and $%
(2m+1,2n+1,2k)$. In this geometry, the equilibrium condition requires $\tau
=-2\sqrt{2}\tilde{\tau}$ , compared to $\tau =-\sqrt{2}\tilde{\tau}$ for \
SQ. Note that it is tension $\tilde{\tau}$ , not $\tau $ which is expected
to be nearly identical for the two competing structures, since it is given
by derivative of the potential, $-\partial _{r}U(r)$ (taken at the distance
of the second nearest neighbor). In contrast, $\tau $ can be considerably
varied by relatively small deformation of the strong $\kappa $--spring.

An obvious choice of unit cell for 2D lattice is a square containing one
atom. For the 3D structure, consider a cube containing eight smaller cubes,
as shown on Figure \ref{SQ}(b). The particles are located at $(2m,2l,2k)+%
\vec{x}$, where $\vec{x}$ is $(1,0,0)$, $(1,0,0)$, $(0,0,1)$, $(1,0,1)$, $%
(1,1,1)$, $(1,1,0)$. The translational symmetry in this structure is
generated by vectors $(2,0,0)$, $(0,2,0)$ and $(1,1,1)$, i. e. the cube on
the picture corresponds to two unit cells. Given the translational symmetry,
the particle at $(1,1,1)$ is equivalent to $(0,0,0)$ (we label them as type
'a' particles), $(0,1,1)$ to $(1,0,0)$ (type 'b') and $(1,0,1)$ to $(0,1,0)$
(type 'c'). We can now obtain Hamiltonian $H$ and calculate the phonon
contribution to free energy.

The number of particles per a unit cell is, $n=1$ \ for SQ and $n=3$ for its
3D-dual. For the 2D phase, displacement of a particle at (m,n) is $\vec{u}%
(m,l)=\sum_{\mathbf{q}}\vec{u}\left( \mathbf{q}\right) e^{i(mq_{1}+nq_{2})}$%
. For the 3D phase, the displacements there are three families of modes,
corresponding to the three types of non-equivalent particles, 'a', 'b', and
'c':

\begin{equation}
\left( 
\begin{array}{c}
\vec{u}(2m+l,2k+l,l) \\ 
\vec{u}(2m+l+1,2k+l,l) \\ 
\vec{u}(2m+l,2k+l+1,l)%
\end{array}%
\right) =\sum\limits_{\mathbf{q}}\left( 
\begin{array}{c}
\vec{u}_{a} \\ 
\vec{u}_{b} \\ 
\vec{u}_{c}%
\end{array}%
\right) _{\mathbf{q}}e^{i(\left( m-l\right) q_{1}+\left( n-l\right)
q_{2}+lq_{3})}
\end{equation}%
The phonon free energy has general form

\begin{equation}
F^{\left( fl\right) }=-k_{B}T\log \int D\left[ u_{i}\left( \mathbf{q}\right) %
\right] \exp \left( -\frac{H}{k_{B}T}\right)  \label{Free-en}
\end{equation}%
Here index $i=1,...,3n$ parameterizes all the modes for a given wave vector $%
\mathbf{q}$, and Hamiltonian $H$ is given by%
\[
H=\sum\limits_{\mathbf{q}}\gamma _{ij}\left( \mathbf{q}\right) u_{i}\left( 
\mathbf{q}\right) u_{j}\left( -\mathbf{q}\right) 
\]%
By performing the Gaussian integration over $u_{i}\left( \mathbf{q}\right) $%
, one can transform Eq. (\ref{Free-en}) into:

\begin{equation}
F^{\left( fl\right) }=\frac{k_{B}TN}{2n}\int \frac{\mathrm{d}\mathbf{q}}{%
\left( 2\pi \right) ^{d}}\log \left( \det \left[ \frac{\hat{\gamma}\left( 
\mathbf{q}\right) }{\pi k_{B}T}\right] \right)  \label{dets}
\end{equation}%
The integration here is performed over a single Brilluen zone, i.e. each
component of the wave vector runs from $-\pi $ to $\pi $.

Note that in our regime ( $\kappa \gg \tilde{\kappa}$ and $\kappa \gg \tau
/a $), one can neglect the coupling between hard and soft modes, which
allows one represent the determinants entering Eq. (\ref{dets}) in
factorized form: $\det \left[ \hat{\gamma}\left( \mathbf{q}\right) \right]
\simeq \det \left[ \hat{\gamma}^{\left( hard\right) }\left( \mathbf{q}%
\right) \right] \det \left[ \hat{\gamma}^{\left( soft\right) }\left( \mathbf{%
q}\right) \right] $. Furthermore, the hard modes have essentially identical
spectra in the both structures (corresponding to one-dimensional chain of $%
\kappa $-springs). Therefore, the difference of the fluctuational free
energies is mainly due to the soft mode contributions. In the case of SQ
phase, there is only one soft mode: 
\begin{equation}
\det \left[ \hat{\gamma}_{2D}^{\left( soft\right) }\left( \mathbf{q}\right) %
\right] =\hat{\gamma}_{zz}\left( \mathbf{q}\right) =\frac{4\sqrt{2}\tilde{%
\tau}}{a}\sin ^{2}\left( \frac{q_{1}}{2}\right) \sin ^{2}\left( \frac{q_{2}}{%
2}\right)
\end{equation}%
For the 3D--dual phase, 3 out of 9 modes are soft, namely $\left(
u_{ax},u_{by},u_{cz}\right) $: 
\begin{equation}
\hat{\gamma}_{3D}^{\left( soft\right) }\left( \mathbf{q}\right) =2\tilde{%
\kappa}\left( 1+3\alpha \right) \delta _{ij}+\frac{\tilde{\kappa}\left(
1+\alpha \right) }{4}\left[ 
\begin{array}{ccc}
0 & \chi \left( q_{1},q_{2}\right) & \chi \left( q_{1},\tilde{q}_{3}\right)
\\ 
\chi \left( q_{1},q_{2}\right) & 0 & \chi \left( q_{2},\tilde{q}_{3}\right)
\\ 
\chi \left( q_{1},\tilde{q}_{3}\right) & \chi \left( q_{2},\tilde{q}%
_{3}\right) & 0%
\end{array}%
\right]
\end{equation}%
here $\alpha =\tilde{\tau}/\left( \sqrt{2}\tilde{\kappa}a\right) $,$\ \ \chi
\left( q,q^{\prime }\right) =\left[ 1+\cos \left( q+q^{\prime }\right) -\cos
\left( q\right) -\cos \left( q^{\prime }\right) \right] $, and $\tilde{q}%
_{3}=2q_{3}-q_{2}-q_{1}$.\ Thus,

\begin{equation}
\det \left[ \hat{\gamma}_{3D}^{\left( soft\right)}%
\right] \simeq 8\tilde{\kappa}^{3}\left( 1+3\alpha \right) ^{3}\left[
1-\varpi_{1}\left( \frac{1+\alpha }{8\left(
1+3\alpha \right) }\right) ^{2}+\varpi_{2}\left( 
\frac{1+\alpha }{8\left( 1+3\alpha \right) }\right) ^{3}\right]
\end{equation}%
Here $\varpi _{1} =\chi ^{2}\left(
q_{1},q_{2}\right) +\chi ^{2}\left( q_{1},\tilde{q}_{3}\right) +\chi
^{2}\left( \tilde{q}_{3},q_{2}\right) $, and $\varpi _{2}=2\chi \left( q_{1},q_{2}\right) \chi \left( q_{1},\tilde{q}%
_{3}\right) \chi \left( \tilde{q}_{3},q_{2}\right) $. Now we can calculate
the fluctuational contribution into the free energy difference between the
two structures:

\begin{equation}
\triangle F^{\left( fl\right) }\equiv F_{2D}^{\left( fl\right)
}-F_{3D}^{\left( fl\right) }\approx -\frac{Nk_{B}T}{2}\left[ \log \left( 
\frac{1}{\alpha }+3\right) +2\log 2-\frac{15}{132}\left( \frac{1+\alpha }{%
1+3\alpha }\right) ^{2}\right]  \label{result}
\end{equation}%
Consistent with the above general arguments, we obtain $F_{2D}^{\left(
fl\right) }-F_{3D}^{\left( fl\right) }<0$, i.e. the entropic effects enhance
the stability of the self--compactified phase. One can identify the physical
origin of each of the three contributions at the above result. The first
logarithmic term is due to an entropic gain of a single fluctuation particle
with all of its neighbors fixed. One can see that these fluctuations are
enhanced in 2D case (for an arbitrary value of parameter $\alpha $), mainly
due to the smaller number of second nearest neighbors there. The second
term, $-Nk_{B}T\log 2$ represent an additional gain due to $q^{4}$
dispersion of the soft acoustic phonon in SQ phase. In contrast, the soft
modes in 3D phase correspond to optical phonons, whose dispersion gives rise
to the negligible third term in Eq. (\ref{result}).

\section{Discussion}

\begin{figure}[tbp]
\includegraphics[
height=1.8741in,
width=4.4278in,
]{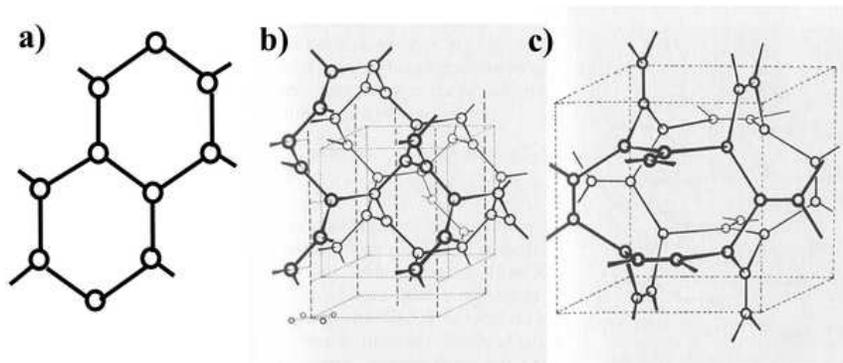}
\caption{2D honeycomb structure(a) and two of its 3D-dual phases (b), taken from Ref. \cite{3Dnets}.}
\label{honey}
\end{figure}

Since the effects responsible for the dominant contributions to our result,
Eq. (\ref{result}), are not specific for the above example, we expect our
conclusions to be rather universal, and applicable to other
self--compactified phases. In fact, the very same binary system introduced
in Ref. \cite{AT}, provides other examples of such phases, e. g. \emph{%
honeycomb} lattice. This structure has several 3D counterparts \cite{3Dnets}, some of which are
shown on Fig.(\ref{honey}). We expect the thermal fluctuations to enhance
the stability of honeycomb with respect to transition to alternative 3D
structures, as in the case of SQ. 

Another aspect of our observation is that the region of the phase diagram
corresponding to quasi-2D structures (like SQ or honeycomb), should expand
when thermal fluctuations are introduced. However, in this study we have not
discuss the transition between solid and liquid phases. In fact, the strong
fluctuations in 2D structures indicate that they should melt earlier than 3D
solid phases, which is consistent with the classical picture. In other
words, the thermal fluctuations enhance the stability of quasi-2D latices
with respect to transition to 3D crystalline structures, but not to the
disordered liquid phase.

Note also that the self-compactified structure need not to be an ideal
crystal, even topologically. Introduction of defects such as disclination or
dislocation, normally associated with 2D melting, is not expected affect our
arguments, as long as the systems remains effectively two-dimensional. That
is because the dominant contribution to the free energy is associated with
length scales of the order of the interparticle distance. Hence, the
macroscopic properties and conformation of these lattices are not relevant
for the discussed phenomenon. As long at the large-scale behavior is
concern, one might expect their properties to be similar to those of
tethered (solid-like) membranes \cite{teth}. It should be noted that the
non-zero bending rigidity is known to stabilize such a membrane with respect
to crumpling transition.

One may give a simple qualitative interpretation to the obtained results.
The self--compactification in zero-fluctuation limit corresponds to the
regime of the interparticle repulsion which is too weak to destabilize the
2D structure, yet strong enough to cause an overall repulsion between two
such layers. Introduction of thermal fluctuations results in an additional
effective repulsion between these 2D layers, an effect very similar to
Helfrich interaction between conventional membranes \cite{membr},\cite%
{helfrich}.

\end{document}